\documentclass[12pt,a4paper]{article}
\usepackage[latin1]{inputenc}
\usepackage[T1]{fontenc}
\usepackage{geometry}
\usepackage{graphicx}
\usepackage{amsmath}
\usepackage{amssymb}
\usepackage{amsthm}
\def\lim{\mathop{\rm lim}\nolimits}

\def\det{\mathop{\rm det}\nolimits}
\def\exp{\mathop{\rm exp}\nolimits}

\def\sup{\mathop{\rm sup}\nolimits}

\def\SL{\mathop{\rm SL}\nolimits}
\def\neigh{\mathop{\rm neigh}\nolimits}

\begin{document}

\centerline{\bf THE SEMI CLASSICAL MAUPERTUIS-JACOBI CORRESPONDENCE:}

\centerline{\bf STABLE AND UNSTABLE SPECTRA}
\bigskip
\centerline{Sergey DOBROKHOTOV ${}^{(*)}$ {\it\&} Michel ROULEUX ${}^{(**)}$}
\bigskip
\centerline {${}^{(*)}$ Institute for Problems in Mechanics of Russian Academy of Sciences}

\centerline {Prosp. Vernadskogo 101-1, Moscow, 119526, Russia, dobr@ipmnet.ru} 

\centerline {${}^{(**)}$ Centre de Physique Th\'eorique and Universit\'e du Sud Toulon-Var, UMR 6207}

\centerline {Campus de Luminy, Case 907, 13288 Marseille Cedex 9, France, rouleux@cpt.univ-mrs.fr}
\bigskip
\noindent {\bf Abstract}: We investigate semi-classical properties of
Maupertuis-Jacobi correspondence for families of 2-D Hamiltonians $(H_\lambda(x,\xi), {\cal H}_\lambda(x,\xi))$, 
when ${\cal H}_\lambda(x,\xi)$ is
the perturbation of a completely integrable Hamiltonian $\widetilde{\cal H}$ veriying some isoenergetic non-degeneracy conditions.
Assuming $\widehat H_\lambda$ has only discrete spectrum near $E$, and
the energy surface $\{\widetilde{\cal H}_0={\cal E}\}$ is separated by some pairwise disjoint Lagrangian tori, we show that most of eigenvalues for 
$\widehat H_\lambda$ near $E$ are asymptotically degenerate as $h\to0$. 
This applies in particular for the determination of trapped modes by an island, in the linear theory of water-waves.
We also consider quasi-modes localized near rational tori. Finally, we discuss breaking of Maupertuis-Jacobi correspondence 
on the equator of Katok sphere.\\

\noindent {\bf 0. Introduction}\\

\noindent {\it 1) Maupertuis-Jacobi correspondence.}\\

KAM theorems for perturbations of an integrable Hamiltonian system provide in phase-space a ``full measure'' set of invariant tori.
For the semi-classical quantization of such a Hamiltonian, we can construct in turn 
quasi-modes associated with these tori, using Maslov canonical operator. 
But the problem of finding even one invariant torus for generic non-integrable Hamiltonian systems is 
in general open; this is achieved in case of Maupertuis-Jacobi
correspondence. Recall the classical~:\\

\noindent {\bf Theorem 1} [Godbillon, Weinstein]: {\it Let $M=M^d$ be a smooth manifold, and ${\cal H}, H\in C^\infty(T^*M)$
two Hamiltonians, sharing a regular energy surface $\Sigma=\{{\cal H}={\cal E}\}=\{H=E\}$. 
Then ${\cal H}, H$ have the same integral curves on $\Sigma$, up to a reparametrization of time.
Hamiltonian vector fields are related by $X_{\cal H}={\cal G}(\tau)X_H$, parametrizations by $dt={\cal G}(\tau)d\tau$.}\\

\noindent {\it Proof}:
Since $\Sigma$ is of codimension one and $\omega$ (the canonical 1-form on $T^*M$) is nondegenerate, 
$V=\{v\in T\Sigma: i_v\omega=0\}$ is one dimensional at each point of $\Sigma$. But $i_{X_{\cal H}}=d{\cal H}=0$ on $\Sigma$,
and similarly $i_{X_H}=dH=0$. Thus $X_{\cal H}$ and $X_H$ are parallel. $\clubsuit$\\

We say that the pair $(H,{\cal H})$ satisfies Maupertuis-Jacobi correspondence at energies $(E,{\cal E})$.\\

\noindent {\it Examples}:\\
\noindent 1)  Geodesic flow and motion in a potential.\\
Let $ds^2=g_{ij}(x)dx^i\otimes dx^j$ be a metric on $M$. Consider
${\cal H}(p,x)={1\over2(E-V)}g^{ij}(x)p_ip_j$, with $E>\sup_M V$, and
$H(p,x)={1\over2}g^{ij}(x)p_ip_j+V(x)$.
Then $(H,{\cal H})$ satisfy Maupertuis- Jacobi correspondence at energies $(E,1)$,
and parametrizations $t$ and $\tau$ are related by $d\tau=(E-V)dt$. 
This was used by Levi-Civita in connexion with Kepler problem. \\

\noindent 2)  Linear water waves theory and Liouville metric on $\Omega\subset{\bf R}^2$.\\
We call Liouville metric on $\Omega$ a conformal metric ${\cal H}(p,x)=g(x)p^2$, such that the geodesic flow is completely integrable with a quadratic
second integral. 
If $\Omega$ is diffeomorphic to an annulus, then ${\cal H}$ is conformally equivalent to a Liouville model metric with additional integral of motion:
$${\cal H}(x,p)=g(x)p^2={p_1^2+p_2^2\over u(x_1)+v(x_2)}, \quad {\cal F}(x,p)={v(x_2)p_1^2-u(x_1)p_2^2\over u(x_1)+v(x_2)}$$
The non-integrable Hamiltonian is the dispersion relation for gravity water waves $H(p,x)=|p|(1+\mu(x) p^2)\tanh(D(x)|p|)$; 
there are one-parameter families $(H,{\cal H}_E)$ and $(H_E,{\cal H})$ satisfying Maupertuis-Jacobi correspondence at energies $(E,1)$,
provided metric $g(x)=g(x,E)$ is conveniently chosen as a function of depth $D(x)$, or vice-versa.\\

\noindent 3) The projective equivalence of metric connexions, e.g. Finsler metrics on the sphere.\\
 
\noindent {\it 2) Maupertuis-Jacobi correspondence for quasi-periodic Hamiltonian flows}.\\

Assume that an integral manifold for ${\cal H}$ consists of the
Lagrangian torus $\Lambda\subset\{{\cal H}={\cal E}\}$ 
such that the Hamiltonian flow $\phi_\tau$ on $\Lambda$ is conjugated to a linear flow
with vector of frequencies 
$\widetilde\omega=(\widetilde\omega_1,\cdots,\widetilde\omega_d)$:
$$({\cal P}(\tau),{\cal X}(\tau))=(P^0(\widetilde\omega\tau+\varphi_0),X^0(\widetilde\omega\tau+\varphi_0)), \quad
\varphi_0\in {\bf T}^d$$
This holds for instance if ${\cal H}$ is completely integrable, or if ${\cal H}$ is
only quasi-integrable and $\Lambda$ is a KAM torus, with Diophantine frequency vector $\widetilde\omega$.\\

\noindent {\bf Theorem 2} [DoRo]: {\it Assume a Diophantine condition on $\widetilde\omega$. Then Maupertuis-Jacobi correspondence 
induces a reparametrization of $\Lambda$ by 
$$\Phi : {\bf T}^d\to{\bf T}^d, \quad \varphi_0\mapsto\varphi=\Phi(\varphi_0)=\varphi_0+\widetilde\omega f(\varphi_0)$$ 
where $f$ is a smooth, periodic function. The corresponding motion on $\Lambda$ is quasi-periodic with frequency vector
$\omega=\widetilde\omega/\langle {\cal G}\rangle$, and~:
$$\det {\partial\Phi\over\partial\varphi_0}={\langle {\cal G}\rangle \over {\cal G}}$$ }\\

In the neighborhood of a Lagrangian immersed manifold in $T^*M$, 
Darboux-Weinstein theorem ensures the existence of a suitable coordinate chart; moreover
in the case of a Lagrangian torus, we can select it of a certain type:\\

\noindent {\bf Theorem 3} [BeDoMa]: {\it Let $\Lambda_0\subset T^*M$ be the Lagrangian torus
$\Lambda_0=\{(p,x)=(P^0(\varphi), X^0(\varphi))\}$.
Then there are action-angle coordinates $(J,\varphi)\in\neigh (I^0;{\bf R}^d)\times{\bf T}^d$,
so that $y'={P(J,\varphi)\choose X(J,\varphi)}$ 
are canonical coordinates defined in a neighborhood of $\Lambda_0$, of the form
$$Y(J,\varphi)=Y^0(\varphi)+Y^J(\varphi)Q(J,\varphi)$$
$\Lambda^J=\Lambda_0+{\cal O}(\iota)$ in the $C^\infty$ topology, $\iota=J-I^0$, 
$J_j={1\over  2\pi}\oint_{\gamma^0_j} P(J,\varphi) d X(J,\varphi)$}\\

These coordinates can be defined without reference to any Hamiltonian system. Now assume $\Lambda_0$ is invariant 
with respect to the flow of Hamilton vector field $X_H$. By the classical Birkhoff normal form (BNF) for $H$, we can find a foliation 
of that neighborhood by almost invariant tori:\\

\noindent {\bf Theorem 4} [Birkhoff]: {\it Let $M$ compact or $M={\bf R}^d$, and
$H\in C^\infty(T^*M)$ (non necessarily integrable); let also
$\Lambda=\Lambda_0\subset T^*M$ be a smooth Lagrangian torus, invariant by the flow of Hamilton vector field $X_H$, conjugated to a linear flow, with
vector of frequencies $\omega$. Then
in the coordinates $(\iota,\varphi)$ above where $\Lambda_0=\{\iota=0\}$, 
$H(p,x)=H_0+\langle\omega,\iota\rangle+{\cal O}(\iota^2)$. 
Moreover if $\omega$ is Diophantine, then
for all $N(\geq2)$ there is a canonical transformation $\kappa_N$,
which maps the zero section onto itself, preserves the action integrals along closed loops, and $H\circ\kappa_N=
H_N(\iota)+{\cal O}(\iota^{N+1})$.}\\

\smallskip
\noindent {{\it 5) Semi-classical quantization} }\\

Consider now a $h$-PDO of order 0 $\widehat H=H^w(x,hD_x;h)$, whose Weyl symbol in
$S^0(M)=\{H\in C^\infty(T^*M): |\partial_x^\alpha\partial_p^\beta H(x,p;h)|\leq C_{\alpha,\beta}\}$ 
has asymptotics $H(x,p,h)\sim H_0(x,p)+hH_1(x,p)+hH_1(x,p)+\ldots, h\to0$
$$\widehat Hu(x,h)=H^w(x,hD_x;h)u(x;h)=(2\pi h)^{-d}\int\int e^{ip(x-y)/h}H({x+y\over2},p;h)u(y)dydp$$
Function $H_0(x,p)$ (or principal symbol of $\widehat H$) is the classical Hamiltonian, and
for simplicity, we shall assume the sub-principal symbol $H_1=0$. We have the following~: \\

\noindent {\bf Theorem 5} [DoRo]: {\it Let $\Lambda=\Lambda_0=\Lambda(I)$ 
be a smooth Lagrangian torus invariant under the Hamiltonian flow of $H_0$, and assume
this flow is conjugated to a linear flow on ${\bf T}^d$, with 
frequency vector $\omega$, $H_0|_{\Lambda}=E$. 

1) Let $0<\delta<1$ and assume the frequency vector $\omega$ be Diophantine. Let $N$ so large that $\delta(N+1)\geq2$, and 
$C_0>0$. Then if $h>0$ is small enough, we can find in a neighborhood of $\Lambda$ 
a (finite) sequence of tori $\Lambda(J',N)$,
almost invariant under $X_H$, modulo ${\cal O}(|\iota|^N)$, 
so that Bohr-Sommerfeld-Maslov (BSM) quantization condition $J'=h\bigl(k+\alpha/4\bigr)$ holds,
provided $J'=J'_k(h)=I+\iota'_k(h)$, for $k\in{\bf Z}^d$,  $|hk-I|\leq C_0h^\delta$. If
\begin{equation}\label{1}
E_k(h)=E+H_N(\iota'_k(h))=E+\langle\omega,\iota'_k(h)\rangle+{\cal O}(|\iota'_k(h)|^{2}), \quad |E_k(h)-E|\leq C_1 h^\delta
\end{equation}
there is a quasi-mode $(u_k(x,h)$, $E_k(h))$ for operator $\widehat H$, i.e.
$(\widehat H-E_k(h)) u_k(x,h)={\cal O}(h^{2})$ in $L^2$ norm. 

2) If $\delta=1$, {\it no irrationality is required} on $\omega$.
Then conclusions of 1) hold for $N=1$, for $k\in{\bf Z}^d$, $|hk-I|\leq C_0h$, when $h>0$ is small enough. 
In particular, the BSM quantization condition writes simply as $J'=h(k+\alpha/4)$. 

3) In both cases, the distance between any $E_k(h)$ and the spectrum of $\widehat H$ is ${\cal O}(h^2)$. 
In particular if the spectrum of operator  $\widehat H$ is discrete near $E$, then there exists an eigenvalue  $\widehat E$ of
$\widehat H$ such that $\widehat E=E_k(h)+{\cal O}(h^2)$.}\\

This theorem makes use essentially of the classical BNF, but as
is well known, classical BNF given in Theorem 4, provided the Diophantine condition, extends as well to all orders in $h$.
In fact, after conjugating  with suitable Fourier integral Operators microlocally near $\Lambda$, 
(which makes use also of a quantization of action-angle variables as given by Darboux-Weinstein theorem) 
$\widehat H$ becomes a $h$-PDO which acts on functions defined on
${\bf T}^d$, and satisfy Floquet periodicity condition
$u(x-k)=e^{2i\pi\langle I+\alpha h/4,k\rangle/h}u(x)$, $k\in {\bf Z}^d$. 
This observation, in the case of high energy asymptotics, follows essentially from [We1] (see also [CdV1] and [Sh]) and
was extended in the semi-classical case, as in [HiSjVu]. So we can improve (5.1) to all orders in $h$,
though it is difficult to obtain the actual perturbation series beyond first order.\\

\noindent {\it 6) Stable and unstable spectra} \\

Let $(\widetilde{\cal H},\widetilde H)$ satisfy Maupertuis-Jacobi correspondence at energies $({\cal E},E)$, with
$\widetilde{\cal H}$ completely integrable; let also ${\cal H}'\in C^\infty(T^*M)$, with $\lambda$
a small coupling constant, 
and ${\cal H}_\lambda=\widetilde{\cal H}+\lambda {\cal H}'$. 

It can happen that there corresponds a smooth
family of Hamiltonians $H_\lambda=\widetilde H(x,\xi)+\lambda H'(x,\xi;\lambda,{\cal E},E)$ such that 
$({\cal H}_\lambda,H_\lambda)$ satisfy Maupertuis-Jacobi correspondence at energies $({\cal E},E)$ for small $\lambda$;
or conversely, given $H_\lambda=\widetilde H+\lambda H'$, that
${\cal H}_\lambda=\widetilde{\cal H}(x,\xi)+\lambda {\cal H}'(x,\xi;\lambda,{\cal E},E)$ and $H_\lambda$
satisfy Maupertuis-Jacobi correspondence at $({\cal E},E)$. 
This is the case in Examples 1 and 2 above.

Now assume the isoenergetic non degeneracy condition (IKAM) on $\widetilde\omega(I)={\partial\widetilde{\cal H}\over\partial I}$ 
holds in $\Sigma$, i.e. 
$$\det 
\begin{pmatrix}{\partial\widetilde\omega(I)\over\partial I}&\widetilde\omega(I)\\ 
{}^t\widetilde\omega(I)&0\end{pmatrix}\neq0$$
which means that $I\mapsto[\widetilde\omega(I)]=(\widetilde\omega_1(I),\cdots,\widetilde\omega_d(I))$ 
restricted to the energy surface $\widetilde{\cal H}={\cal E}$ is a (local) isomorphism on the projective space.
When $\sigma>d-1$, and for $c>0$ small enough, we define a KAM set on $\Sigma$, as the Cantor set~:
$$K_{c,\sigma}=\{I\in\Sigma=\widetilde{\cal H}^{-1}({\cal E}): \forall k\in{\bf Z}^d\setminus0, \ |\langle k,\widetilde\omega(I)\rangle|
\geq c|k|^{-\sigma}\}$$
whose complement has a small measure (of order $c$) as $c\to0$.

For $I\in K_{c,\sigma}$, we know that the KAM torus $\Lambda(I)$ survives small perturbations ${\cal H}_\lambda=\widetilde{\cal H}+\lambda{\cal H}'$ 
of $\widetilde{\cal H}$, and
the Hamiltonian flow for ${\cal H}_\lambda$ is again quasi-periodic on a deformation
$\Lambda_\lambda(I)$ of $\Lambda(I)$ with a frequency vector proportional to $\widetilde\omega(I)$ (see [Bo,Theorem 1.2.2]).

Consider now $h$-PDO $\widehat H_\lambda$ with principal symbol $H_0(x,\xi)=H_\lambda(x,\xi)$. 
By Theorem 5 we can construct a quasi-mode for $\widehat H_\lambda$, corresponding to 
asymptotic eigenvalues $E_k(h)$, for all lattice points $kh$ (possibly shifted by Maslov index), within a distance of any KAM set on $\Sigma$,
not exceeding $h^\delta$. The dimension of the span of the corresponding asymptotic eigenfunctions $\varphi_k(h)$ is about
$(2\pi h)^{-d}|K_{c,\sigma}|$. 
This set of lattice points has several ``connected components'', which consist in
the {\it stable spectrum} induced by Maupertuis-Jacobi correspondence, and are separated by so-called ``resonance'' zones,
the {\it unstable spectrum}. The unstable spectrum is associated
with quasi-modes concentrated on rational Lagrangian tori, or elliptic periodic orbits, or with 
so-called Shnirelman quasi-modes concentrated on connected components of $\Sigma$ between KAM tori.\\

\noindent {\it 7) Isoenergetic KAM Symmetries and degeneration of the spectrum.} \\

Our main result is the semiclassical counterpart of a theorem by Shnirelman.
Given some $J(h)\subset{\bf N}$, with $|J(h))|\to\infty$ as $h\to0$, and
$J'(h)\subset J(h)$, we say that is of relative density 1 iff 
$\lim_{h\to0}{|J'(h)|\over|J(h))|}=1$. We say also that $\widehat H$ is time-reversal invariant iff 
$\widehat H\Gamma=\Gamma\widehat H$, where $\Gamma(x,\xi)=(x,-\xi)$. We have:\\

\noindent{\bf Theorem 6}: {\it Assume $d=2$. Let $H\in S^0(M)$, such that $\widehat H$  has discrete spectrum in 
$I(h)=[E-h^\delta,E+h^\delta]$, $0<\delta<1$. 
Let
$J(h)=\{j\in {\bf N}: \lambda_j(h)\in I(h)\}$ label the eigenvalues in $I(h)$, counted with multiplicity.
On the other hand, let ${\cal H}$ be completely integrable, or a small perturbation 
${\cal H}=\widetilde{\cal H}+\lambda{\cal H}'$ of an integrable Hamiltonian $\widetilde{\cal H}$, satisfying IKAM condition.
Assume that $({\cal H},H)$ satisfy Maupertuis-Jacobi correspondence at energies $({\cal E}, E)$, and 
$\Sigma=\{H=E\}=\{{\cal H}={\cal E}\}$ is compact, diffeomorphic to  
${\bf T}^2\times{\bf S}^1 \ \hbox{or} \  {\bf T}^2\times[0,1]$
so that it is separated by any 2 pairs of invariant tori. 
Assume also $\widehat H$ is time-reversal invariant, and there are 4 invariant tori (separated in phase-space) with Diophantine frequency vectors
$$\Lambda_1, \ \Lambda_2=\Gamma(\Lambda_1), \ \Lambda_3, \ \Lambda_4=\Gamma(\Lambda_3)$$
Then there exists $J'(h)\subset J(h)$ of relative density 1 such that 
$\forall j\in J(h): |\lambda_{j\pm1}(h)-\lambda_j(h)|={\cal O}(h^\infty)$.}\\

It holds in particular in Example 2.
Thus, most of the spectrum of $H$ in a $h^\delta$-neighborhood of $E$ is asymptotically degenerated, due to the time-reversal invariance,
and existence of invariant tori separated in phase-space. This {\it phase-space tunneling} is given a precise form in case of the 
Liouville metric ${\cal H}$ [DoSh], where the splitting between semi-classical eigenvalues turns out to be exponentially small. Thus 
phase-space tunneling reflects also in Maupertuis-Jacobi correspondence.

The proof of this theorem is close to this of [Sh] and consists in 3 steps: 1) 
The construction of a ``quasi-projector'' quantizing a ``smeared'' characteristic function of the set
$\Sigma_3$ between $\Lambda_1$ and $\Lambda_2$, that depends essentially on the action variables $\iota$ 
near $\Lambda_j$, given by the BNF; and similarly with $\Lambda_3$ and $\Lambda_4$. 2) The use of semi-classical measures
corresponding to the spectrum of $\widehat H$ in a $h^\delta$-neighborhood of $\Sigma=\Sigma_E$, that converge weakly to the normalized
Liouville measure on $\Sigma_E$.
3) A dichotomy argument that sorts out measures supported on different splittings of $\Sigma_E$ by the $\Lambda_j$'s.

The conclusion is somewhat weaker than the corresponding one in [Sh] for high energy asymptotics of Laplace-Beltrami
eigenfunctions on $M$, since it only gives information relative to a subsequence (of density 1) of eigenvalues.\\

\noindent {\it 8) Quasi-modes for rational Lagrangian tori and Larmor precession}.\\

We consider next the situation where
the frequency vector associated with the flow of $X_{\cal H}$ on $\Lambda$ is rational (then we say for short that $X_{\cal H}$ has rational flow).
It may happen that the flow of $X_H$ on $\Lambda$ induced by Maupertuis-Jacobi correspondence is again rational~; in this case we 
can construct by means of Theorem 5 quasi-modes for $\widehat H$ in a $h^{1/2}$-nghbd of $\Lambda$. 
In general however, the flow of $X_H$ on $\Lambda$ is again conjugated to a linear flow, 
but with a frequency that generally depends on the initial condition. 

Making a linear transformation $T\in{\SL}_2({\bf Z})$ on the angles, we can assume that
$\widetilde\omega=(\widetilde\omega_1,0)$. Maupertuis-Jacobi correspondence induces a reparametrization of time, of the 
form $dt={\cal G}(\varphi_1+\widetilde\omega_1\tau,\varphi_2)d\tau$. We have 
set $\varphi=(\varphi_1,\varphi_2)\in{\bf T}^2$ (for simplicity, we restrict to $d=2$.)
There is a reparametrization of ${\bf T}^2$, of the form 
$$\Phi : \varphi\mapsto\psi=\Phi(\varphi)=\varphi+\widetilde\omega g(\varphi)$$ 
where $g$ is a smooth, scalar periodic function.
The motion on $\Lambda$ induced by this reparametrization is periodic with frequency vector $\omega=(\omega_1,\omega_2)=(\omega_1,0)$, 
where $\omega_1=\omega_1(\psi_2)={\widetilde\omega_1\over\langle {\cal G}\rangle_{\psi_2}}$, and
$\langle {\cal G}\rangle_{\psi_2}$ denotes the average with respect to $\psi_1$. It is a smooth periodic
function of $\psi_2=\varphi_2\in{\bf T}$. 
Near $\Lambda$ we apply Darboux-Weinstein theorem in the special form given in Theorem 3, we can write 
Hamiltonian $H$ in some symplectic coordinates $(x,\xi)$ where $x=(x_1,x_2)\in{\bf T}^2$ 
stands for $\psi$ above, $\xi=(\xi_1,\xi_2)\in{\bf R}_+^2$ the dual coordinate and $\Lambda$ is given 
by $\xi=0$. With $H_0=H|_\Lambda$, this gives
$H=H_0+\omega_1(x_2)\xi_1+a(x,\xi)$
where $a(x,\xi)={\cal O}(\xi^2)$ is a smooth periodic function on $x\in{\bf T}^2$.

There is no canonical way to determine the motion on nearby 
tori $\Lambda^J$, but 
under some ellipticity condition, this motion can be identified with a Larmor precession in a varying magnetic field,
see [ArKoNe,Sect.6.4]. 
In this model the  ``fast variable'' $x_1$ stands for the direction of the ``unperturbed
orbits'' (the small circles), and the ``slow variable'' $x_2$ for the direction of the ``drift''. Averaging in
variable $x_1$, for any $N$ we can reduce $H$ to 
$H(x_2,\xi)=H_0+\omega_1(x_2)\xi_1+H_N(x_2,\xi)+{\cal O}(|\xi|^{N+1})$
where $H_N(x_2,\xi)$ is a polynomial in $\xi$ of degree $N$ and vanishing of order 2 at $\xi=0$. We notice that 
$H'_N(x_2,\xi)=\omega_1(x_2)\xi_1+H_N(x_2,\xi)$ is in involution with $\xi_1$, so we are reduced to a 1-D problem, and discuss
according to the components of the Reeb graph. A simple model is given by $H_N(x_2,\xi)={1\over2}\xi_2^2$, 
and the spectrum is obtained, in first approximation, by considering the family of Schr\"odinger operators
$P(x_2,hD_{x_2})={1\over2}(hD_{x_2})^2+\omega_1(x_2)k_1$, $k_1\in2\pi h{\bf Z}$, $|k_1|\leq h^\delta$. \\

\noindent{\it 9) Aharonov-Bohm effect on the sphere and projectively equivalent Finsler structures}.\\

We examine here the {\it breaking} of Maupertuis-Jacobi correspondence in the following example:
Consider the Lagrangian on the unit sphere $M={\bf S}^2$ of ${\bf R}^3$
\begin{equation}\label{2}L(x,v)=\sqrt{{1\over2}\vec v^2}+\langle\vec A(x),\vec v\rangle
\end{equation}
where $\vec A(x)=\alpha(-{\sin q_1\over\cos q_2}, {\cos q_1\over\cos q_2}, 0)$ 
in equatorial coordinates $(q_1,q_2)$; this
is a radially symmetric potential vector, inducing a closed form $d\vec A(x)=0$, but not exact, due to
the singularity of $A(x)$ at the poles. We call $L$ the ``Aharonov-Bohm Lagrangian'' on ${\bf S}^2$. 
When $|\alpha|<1$, (2) defines a Finsler metric on $M={\bf S}^2$, which is not
{\it reversible}, due to the linear term.
Finsler metrics we consider here are a special class known as ``Randers metrics'', i.e. metrics of the form
\begin{equation}F(x,v)=\sqrt{g_x(v,v)}+g_x(v,X)\end{equation}
where $g_x$ is a Riemannian metric tensor on $M$ and $X$ a real vector field satisfying $g_x(X,X)<1$. 
With a Finsler metric, we associate a Hamiltonian by the usual prescription. In the case of 
a Randers metric, this yields a ``Randers symbol'' on $T^*M$, having the form~:
\begin{equation}H(x,\xi)=\sqrt{\widetilde h_x(\xi,\xi)}+\langle Y,\xi\rangle=\lambda(x,\xi)+\eta(x,\xi)\end{equation}
where $\widetilde h_x$ is a positive definite quadratic form on $T_x^*M$ and $Y$ a real vector field on $M$, satisfying 
$\langle Y,\xi\rangle=\widetilde h_x(Y,\xi)$, 
$\widetilde h_x(Y,Y)<1$. 
We can identify Finsler metric (2) with the famous Katok metric on ${\bf S}^2$ constructed as follows. Let 
$g$ be the standard metric tensor 
on ${\bf S}^2$, and $Y_0\in T{\bf S}^2\setminus0$ is the generator of a group of rotations
$R_0(t)$ of period $2\pi$.
We take $Y=\alpha Y_0$, $\alpha\in]-1,1[$, so that $g_x(Y,Y)<1$. So Katok Lagrangian identifies with Aharonov-Bohm Lagrangian if we take
$Y=\vec A$.

The geometry of Katok sphere is well-understood, see [Tay] and [Zi].
In particular, when $\alpha$ is rational, the flow $\exp tH_\eta(x,\xi)$ is completely periodic,
as is the geodesic flow on the standard sphere ${\bf S}^2$.
For irrational $\alpha$ instead, there are only 2 closed geodesics $\gamma_\pm$, both supported on the equator $\gamma$, but swept with different speeds
$1\pm\alpha$, due to the fact that the metric is not reversible.

If $\vec A$ were smooth everywhere, the 2 metrics (standard and Katok) would be
{\it projectively equivalent}, in the sense they have same geodesics; this is broken by the singularity of $\vec A$ at the poles.

In fact, projective equivalence is a special case of Maupertuis-Jacobi correspondence in case of Finsler metrics. 
Standard and Katok metrics on ${\bf R}^2$ fulfill only partially this correspondence at energies $(1,1)$, in the sense that their 
only common Hamiltonian orbit is the (lift of) equator $\gamma$. 
We know from [Zi], that if $\alpha$ is irrational, then
the Poincar\'e maps of the closed geodesics $\gamma_\pm$ are tangent to rotations with angle 
${2\pi\over1\pm\alpha}$ respectively. In particular, $\gamma_\pm$ are
of elliptic type with irrational exponents.
This allows to construct, by the method of complex germs,  quasi-modes supported on $\gamma_\pm$, see e.g. [BrDoSe-ZeTu].\\

\noindent {\it References}\\

\noindent [ArKoNe]  V.Arnold, V.Kozlov, A.Neishtadt. Mathematical aspects of classical and celestial mechanics. Encyclopaedia of Math. Sci.,
Dynamical Systems III, Springer, 2006.

\noindent [BaWe] S.Bates, A.Weinstein. Lectures on the geometry of quantization. Berkeley Math. Lect. Notes 88,
American Math. Soc. 1997.

\noindent [BeDoMa] V.Belov, S.Dobrokhotov, V.Maksimov. Explicit formulas for  generalized action-angle variables
in a neighborhood  of an isotropic torus and their applications. Theor. Math. Phys., 135(3),p.765-791, 2003.

\noindent [Bo] J.B. Bost. Tores invariants des syst\`emes dynamiques hamiltoniens. S\'eminaire Bourbaki, Expos\'e 639, Vol. 1984-85.

\noindent [BrDoSe-ZeTu] J.Br\"uning, S.Dobrokhotov, S.Sekerzh-Zenk'ovich, T.Tudorovskiy. Spectral series of Schr\"o- dinger operators
in a thin wave-guide with periodic structure 2. Russian J. Math. Phys. 18(1), p.19- ,2011.

\noindent [CdV]  Y.Colin de Verdi\`ere. {\bf 1}. Modes et quasi-modes sur les vari\'et\'es riemanniennes. Inventiones Math.
43, p.15-52, 1977. {\bf 2} M\'ethode de moyennisation en M\'ecanique
semi-classsique. Journ\'ees Equations aux D\'eriv\'ees partielles, Expos\'e No 5, Saint Jean de Monts, 1996.

\noindent [DoRo] S.Dobrokhotov, M.Rouleux. 
The semi-classical Maupertuis-Jacobi correspondence for quasi-periodic hamiltonian flows with applications to linear 
water waves theory. Asymptotic Analysis 74, p.33-73, 2011.

\noindent [DoZh] S.Dobrokhotov, A.Shafarevich. ``Momentum'' tunneling between tori and the splitting of 
eigenvalues of the Laplace-Beltrami operator on Liouville surfaces. Math. Phys. Anal. Geometry 2, p.141-177, 1999. 

\noindent [HiSjVu] M.Hitrik, J.Sj\"ostrand, S.Vu-Ngoc. Diophantine tori and spectral asymptotics for non-self adjoint 
operators. Amer. J. Math.  129(1), p.105-182, 2007. 

\noindent [Laz] V.Lazutkin. KAM theory and semi-classical approximation to eigenfunctions. Springer, 1993.

\noindent [Sh] A.V.Shnirelman. On the asymptotic proporties of the eigenfunctions in the region of chaotic motion.
Addendum to [Laz].

\noindent [Ta] M.Taylor. Finsler structures and wave propagation, {\it in}: V.Isakov (ed.), Sobolev spaces in Mathematics III.
Int. Math. Series, Springer, 2009.

\noindent [We] A.Weinstein. On Maslov quantization condition, {\it in}:  Fourier Integral Operators and 
Partial Differential Equations. J.Chazarain, ed. Lecture Notes in Math. 469, Springer, p.361-372, 1974. 

\noindent [Zi] W.Ziller. Geometry of the Katok examples. Ergod. Theor. Dyn. Sys. 3, p.135-157, 1982.

\end{document}